# Benefits of InterSite Pre-Processing and Clustering Methods in E-Commerce Domain

Sergiu Chelcea[1], Alzennyr Da Silva[1&2], Yves Lechevallier[2], Doru Tanasa[1], Brigitte Trousse[1]

[1] AxIS, INRIA Sophia-Antipolis
2004, Route des Lucioles, B.P. 93
06902 Sophia Antipolis Cedex, France
`{Sergiu.Chelcea,Doru.Tanasa,Brigitte.Trousse}`@inria.fr
http://www-sop.inria.fr/axis/
[2] AxIS, INRIA Rocquencourt
Domaine de Voluceau, Rocquencourt, B.P. 105
78153 Le Chesnay Cedex, France
`{Alzennyr.Da_Silva,Yves.Lechevallier}`@inria.fr
http://www-rocq.inria.fr/axis/

**Abstract.** This paper presents our preprocessing and clustering analysis on the clickstream dataset proposed for the ECML\PKDD 2005 Discovery Challenge. The main contributions of this article are double. First, after presenting the clickstream dataset, we show how we build a rich data warehouse based an advanced preprocesing. We take into account the intersite aspects in the given e-commerce domain, which offers an interesting data structuration. A preliminary statistical analysis based on time period clickstreams is given, emphasing the importance of intersite user visits in such a context. Secondly, we describe our crossed-clustering method which is applied on data generated from our data warehouse. Our preliminary results are interesting and promising illustrating the benefits of our WUM methods, even if more investigations are needed on the same dataset.

## 1 Introduction

The daily access of an Internet Web site can today easily rise to a number of access in millions of pages, executed by a large amount of users spread all over the world. Web usage mining [9] is the application of data mining technologies on large logs files, collected from Web servers accesses. Examples of such applications include: improvements of web sites design, system performance analyses as well as network communications, understanding user reaction and motivation, automated clustering and building adaptive web sites [4][7][12].

E-commerce organizations are especially interested in the insight that web usage mining provides [5][8]. Such insight helps not only to improve their web site, but also their services and marketing strategies (promotions, banners, etc.).

This paper aims at analyzing the clickstream dataset provided for the Challenge PKDD 2005. According to different time periods, we aim at discovering the usage and the server charge in the context of multiple Web sites in the e-commerce domain. This paper is organized as follows. Section 2 presents the analyzed dataset. Then section 3 presents our Intersite advanced data preprocessing. Next, before concluding in Section 5, Section 4 describes our clustering approach and the respective analysis.

## 2 Used Clickstream Dataset Description

The proposed clickstream dataset consists in 576 log files with a total of 3,617,171 requests for page views. These requests were made on seven different e-commerce Web sites from the Czech Republic (see Table 2). Each log file contains all the requests recorded during one hour on the seven e-commerce Web sites, whilst all the files cover a continuous 24 days period starting from 09:00AM 20$^{th}$ January 2004 until 08:59AM on the 13$^{th}$ February 2004.

The log files were in csv format, with each line containing a request for a page on one of the 7 e-commerce Web servers and having the 6 following fields (Table 1):
- *ShopID*: an ID of the e-commerce Web server (also denoted as shop) that received the request;
- *Date*: the Unix time of the request (seconds since 00:00:00 1$^{st}$ January 1970);
- *IP address*: the computer's IP address of the user making the request;
- *SessionID*: a php session id automatically generated for each new *visit* on each server (unique IDs);
- *Page*: the requested resource (page) on the server;
- *Referrer*: the referrer of the requested page.

**Table 1.** Format of page requests

| ShopID | Date | IP address | SessionID | Page | Referrer |
|---|---|---|---|---|---|
| 11 | 1074585663 | 213.151.91.186 | 939dad92c4…84208dca | / | |
| 11 | 1074585670 | 213.151.91.186 | 87ee02ddcff…7655bb9e | /ct/?c=148 | http://www.shop2.cz |

A table containing the seven shops names and their ID was provided (see Table 2, first two columns). For confidentiality reasons, the names of the seven e-commerce Web sites have been anonymized in this table as well as in the *Referrer* field when present.

**Table 2.** Number of requests per shop

| ShopID | Site name (shop) | #Requests |
|---|---|---|
| 10 | www.shop1.cz | 509,688 |
| 11 | www.shop2.cz | 400,045 |
| 12 | www.shop3.cz | 645,724 |
| 14 | www.shop4.cz | 1,290,870 |
| 15 | www.shop5.cz | 308,367 |
| 16 | www.shop6.cz | 298,030 |
| 17 | www.shop7.cz | 164,447 |

The Web pages on these servers are interconnected, meaning users can navigate from one *shop* to another using only the links in the pages (clicks). However, since the *SessionID* is generated when first entering a page of a web shop, a user will get a new *SessionID* when he/she switches to a yet unvisited shop.

The *Page* field contains the path on the server corresponding to the *ShopID* field to the requested page. There are 21 "page types" corresponding to the distinct 21 first level syntactic topics of all pages (*ct, ls ,dt, znacka, akce, df ,findf, findp, setp, poradna, kosik, obchody-elektro, kontakt, faq, onakupu, splatky, mailc, mailp, mailf, mailr*). Using these page types and their provided descriptions, we can thus find out if the user has made a request for a specific product, category or theme, if some filters were applied, etc. For example, in the second entry presented in Table 1, the user has requested the products from the Earphones category (code 148).

To describe the variables present in the requested pages, another four tables were made available: *kategorie* (product categories), *list* (products), *znacka* (products brands) and *tema* (product themes).

## 3  Advanced Intersite Pre-processing

In order to prepare the dataset for our analysis, we used a recently proposed methodology for multisites logs data preprocessing [10][11], which extends the Cooley's previous work [1].

Unfortunately, the provided raw data was not formatted in the CLF (Common Log Format [14] and some fields were not available (i.e. the status code, the user agent, logname). Thus, we rely only on the *SessionID* to identify a user's visit.

The data preprocessing was done in four steps: *data fusion*, *data cleaning*, *data structuration*, *data summarization*.

Generally, in the *data fusion* step, the log files from different Web servers are merged into a single log file. This was already done, so we only merged the 579 log files. We also changed the *Date* format into Gregorian time in order to facilitate our analysis interpretation, and merged the *Page* and the *ShopID* fields into the *URL* field (see Table 3) in order to have same format as the *Referrer* field.

**Table 3.** Transformed log lines

| Datetime | IP | SessionID | URL | Referrer |
|---|---|---|---|---|
| 2004-01-20 09:01:03 | 213.151.91.186 | 939dad92c4…84208dca | http://www.shop2.cz/ | - |
| 2004-01-20 09:01:10 | 213.151.91.186 | 87ee02ddcff…7655bb9e | http://www.shop2.cz/ct/?c=148 | http://www.shop2.cz/ |

During the data cleaning, the non-relevant resources are eliminated (e.g. jpg, js files). Here, we assume that all requests have succeeded (code 200). On this dataset, a user changing shops can have during a single visit multiple *SessionIDs*, one on each shop. For this reason, we decided to group such *SessionIDs* that belong to a single user (same IP) into a *Group of SessionIDs*, corresponding to the user's actual visit. This was done by comparing the *Referrer* with the *URLs* previously accessed (in a reasonable time window), each time the user moves to another shop. If the *Referrer* (actually a page on another shop) was previously accessed, we group the two *SessionIDs* together (the actual one and the one on the previous shop). We thus grouped the existing 522,410 *SessionIDs* into 397,629 groups, equivalent to a 23.88% reduction in the user visits number. For example, in Table 3 we grouped into the same session group the two *SessionIDs* because the *Referrer* of the second line was recently requested from the same IP address and recorded in the URL field of the first line. Thus, we obtained cross-server user visits which can be used to perform global analyses on all the shops.

Fig. 1a (global visits) and 1b (multi-shop visits) show clearly the low number of customers new visits on Saturdays and Sundays during the lunch time and the high number on Tuesdays and Wednesdays.

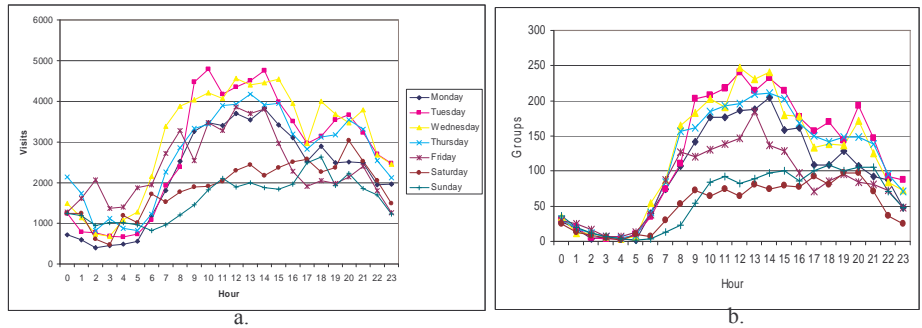

**Fig. 1.** Visits per days and hours: (a) globally, (b) multi-shop

Finally, after identifying each variable in the accessed URL and their corresponding descriptions, we defined an extendable relational database model to use, model inspired from [10]. In the next section, we present our crossed clustering method and we illustrate the usefulness of the structured pre-processed data and the flexibility of our DB model. We decided to use as time units slices of date and hour, which we call *Time Periods* for our crossed clustering approach.

## 4  Crossed Clustering Approach for Time Periods/Product Data Analysis

In the crossed clustering approach [6] [13], as in the classical clustering algorithm, the criterion optimized is based on the best fitting between classes of objects and their representation. Some authors [2] [3] proposed the maximization of the chi-squared criterion between rows and columns of a contingency table.

For our analysis we added a crossed table to the relational DB model. Each line describes an individual that covers the couple weekday and hour of requests for the *ls* pages in shop 4 (the more visited one according to Table 2). The column describes one multi-categorical variable which represents the number of products requested by users into a specific time slice (see Table 4, where we have 7 x 24 individuals).

**Table 4.** Quantity of products requested by weekday x hour and registered on shop 4

| Weekday x Hour | Product (number of requests) |
|---|---|
| Monday_0 | Built-in electric hobs (10), Built-in dish washers 60cm (64), Corner single sinks (50), ... |
| Monday_1 | Free standing combi refrigerators (44), Corner single sinks (50), Built-in hoods (60), ... |
| ... | ... |
| Sunday_22 | Built-in microwave ovens (27), Built-in dish washers 45cm (38), Built-in dish washers 60cm (85), ... |
| Sunday_23 | Built-in freezers (56), Kitchen taps with shower (45), Garbage disposers (32), ... |

The 168 periods of time summarize 490,883 requests on all products from shop 4. Table 5 presents the results after applying the crossed clustering method [6] specifying 7 classes of periods and 5 classes of products.

**Table 5.** Confusion table

|  | Product_1 | Product_2 | Product_3 | Product_4 | Product_5 | Total |
|---|---|---|---|---|---|---|
| Period_1 | 2847 | 5084 | 3284 | 2265 | 2471 | 15951 |
| Period_2 | 11305 | 31492 | 12951 | 1895 | 9610 | 67253 |
| Period_3 | 33107 | 55652 | 36699 | 5345 | 20370 | 151173 |
| Period_4 | 22682 | 46322 | 30200 | 5165 | 27659 | 132028 |
| Period_5 | 9576 | 20477 | 19721 | 2339 | 7551 | 59664 |
| Period_6 | 1783 | 3515 | 2549 | 392 | *11240* | **19479** |
| Period_7 | 15019 | 14297 | 8608 | 1397 | 6014 | 45335 |
| Total | 96319 | 176839 | 114012 | 18798 | **84915** | 490883 |

Surprisingly the product class 5 was defined only by one product, namely *Free standing combi refrigerators,* consulted predominantly on Fridays from 17:00 to 20:00 (period 6). It is important to note that although this product belongs to a class product responsible only for 17.3 % of the total requests on shop 4, the requests occurring on its period is 57.7 % based on this product. In other words it means this product is the most requested one on this weekday and time period. Such information could be used on marketing strategies, like cross selling, fast promotions, etc. More information on the seven Web sites content and structure could have been very useful to raise interesting problems and to facilitate the results interpretation from the business point of view.

## 5   Conclusions and Future Works

In this paper, we first proposed a pre-processing method in the context of a multi-shop e-commerce domain. Our analysis on the proposed PKDD Web logs showed the great flexibility offered by the built data warehouse and the benefits of an enriched data structuration in terms of customer visits. Secondly we presented an original clustering analysis based on an efficient clustering method applied on Web time period-based clickstreams: our first results on the used dataset in a short time due to the PKDD challenge are promising. Such an analysis allows us to identify best hours for marketing strategies, like fast promotions, on-line advices and publish banners, etc. Others analyses could be planned in the future, exploiting for example the link between the consumer activities and the time periods by shop or focusing on multi-shop user visits, etc.